\documentclass[aps,prl,twocolumn,superscriptaddress, 10pt]{revtex4-1} 

\usepackage{graphicx} 
\usepackage{dcolumn}
\usepackage{bm}
\usepackage{amsmath,latexsym,tabularx}
\usepackage{multirow}
\usepackage[export]{adjustbox}
\usepackage{hyperref}
\usepackage{upgreek}
\hypersetup{
  colorlinks=true,
  urlcolor=blue,
  linkcolor=blue,
  citecolor=blue
}
\usepackage[inkscapelatex=false]{svg}

\newcommand{\affilUMass}[0]{Department of Electrical and Computer Engineering, University of Massachusetts Amherst, Amherst, MA 01003}
\newcommand{\affilUCSB}[0]{Department of Electrical and Computer Engineering, University of California Santa Barbara, Santa Barbara, CA 93106}

\newcommand{\Sr}{$^{88}$Sr$^+$}
\newcommand{\SiN}{$\text{Si}_3 \text{N}_4~$}

\begin{document}

\title{Trapped ion qubit and clock operations with a visible wavelength photonic coil resonator stabilized integrated Brillouin laser}

\author{Nitesh Chauhan$^1{}^*$, Christopher Caron$^2{}^*$, Jiawei Wang$^1$, Andrei Isichenko$^1$, Nishat Helaly$^2$, Kaikai Liu$^1$, Robert J. Niffenegger$^2{}^{\dag}$, Daniel J. Blumenthal$^1{}^{\dag}$
\\ \normalsize $^{1}$\affilUCSB
\\ \normalsize $^{2}$\affilUMass
\\ \normalsize Corresponding Authors:$^{\dag}$rniffenegger@umass.edu, danb@ucsb.edu
\\ \normalsize $^*$These authors contributed equally \\ }

\date{\today}

\begin{abstract}

Integrating precise, stable, ultra-low noise visible light lasers into atomic systems is critical for advancing quantum information sciences and improving scalability and portability. Trapped ions are a leading approach for high-fidelity quantum computing, high-accuracy optical clocks, and precision quantum sensors. However, current ion-based systems rely on bulky, lab-scale precision lasers and optical stabilization cavities for optical clock and qubit operations, constraining the size, weight, scalability, and portability of atomic systems. Chip-scale integration of ultra-low noise lasers and reference cavities operating directly at optical clock transitions and capable of qubit and clock operations will represent a major transformation in atom and trapped ion-based quantum technologies. However, this goal has remained elusive. Here we report the first demonstration of chip-scale optical clock and qubit operations on a trapped ion using a photonic integrated direct-drive visible wavelength Brillouin laser stabilized to an integrated 3-meter coil-resonator reference cavity and the optical clock transition of a \Sr ion trapped on a surface electrode chip. We also demonstrate for the first time, to the best of our knowledge, trapped-ion spectroscopy and qubit operations such as Rabi oscillations and high fidelity (99\%) qubit state preparation and measurement (SPAM) using direct drive integrated photonic technologies without bulk optic stabilization cavities or second harmonic generation. Our chip-scale stabilized Brillouin laser exhibits a 6 kHz linewidth with the 0.4 Hz quadrupole transition of \Sr and a self-consistent coherence time of 60 $\mu$s via Ramsey interferometry on the trapped ion qubit. Furthermore, we demonstrate the stability of the locked Brillouin laser to 5$\times10^{-13}/ \sqrt{\tau}$ at 1 second using dual optical clocks.
These results, facilitated by the low thermal-refractive noise 3-meter coil-resonator cavity and the ultra-low loss CMOS foundry compatible silicon nitride integration platform, pave a direct path towards monolithic integration of stabilized lasers within trapped ion chips. This advancement toward development of portable trapped-ion quantum sensors and clocks, heralds an era of portable quantum technologies.

\end{abstract}

\maketitle

\section{Introduction}
\vspace{-0.2in}
Trapped ions are a fundamental technology platform for quantum information sciences, enabling quantum information processors \cite{bruzewicz2019trapped}, optical clocks \cite{ludlow2015optical}, precision quantum-logic spectroscopy \cite{schmidt2005spectroscopy}, and quantum sensors \cite{ruster2017entanglement}. Experiments based on ion trap technology employ a physics package that contains the ion trap and external laser and optical infrastructure used in the process of ion cooling, manipulation, interrogation, and readout. Today's optical infrastructure occupies table tops and racks of equipment, and supports functions including light generation and stabilization, beam control and delivery to the ion, modulation, detection, and control, with light delivered to the trapped ion using optical fiber and free-space optics. Not only does the volume and size of current optical infrastructure add to the size, weight and cost of trapped-ion systems, but it also limits reliability, scalability, and even experimental precision and performance, particularly through the generation and accumulation of laser phase noise as light reaches the ion. There has been great success in miniaturizing the trapped ion physics package, and already trapped ion qubit systems are some of the smallest quantum computing systems in the world \cite{pogorelov2021compact}. However, the  miniaturization through integration of the remaining system volume comprised of precision optical laser systems and moving these components closer to the trap itself has yet to be demonstrated.

\begin{figure*}[]
\includegraphics[width=0.97\textwidth]{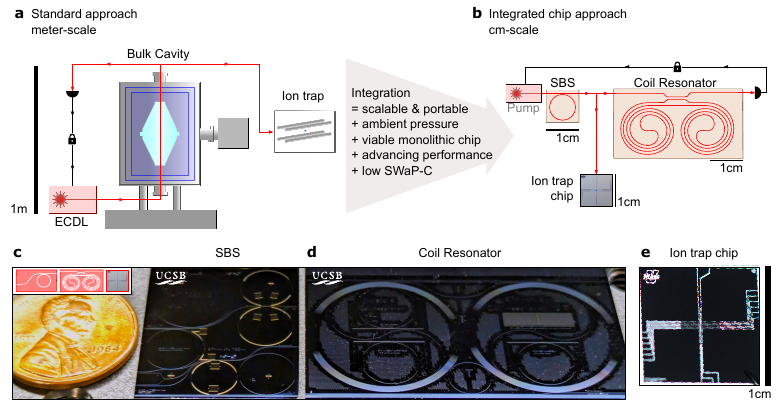}
\caption{
\textbf{Chip-scale laser stabilization for portable optical clocks and trapped ion qubits: }
\textbf{a},
Conventional optical clocks utilize carefully isolated bulk cavities to narrow and stabilize an external cavity diode (ECDL) laser for interrogating optical clock transitions. 
\textbf{b},
We demonstrate a chip-scale integrated approach for generating and stabilizing the ultra-narrow linewidth laser. 
We use a photonic integrated direct-drive 674 nm ultra-low phase noise Stimulated Brillouin Scattering (SBS) laser which is stabilized to a photonic integrated 3-meter coil resonator reference cavity. 
Picture of the integrated SBS chip (\textbf{c}) and of the integrated coil resonator chip (\textbf{d}) with a penny for scale.
\textbf{e},
Picture of the integrated surface electrode ion trap.
}
\label{fig:vision}
\end{figure*}

Recently, many other functions necessary for ion trap operation have been monolithically integrated into the surface electrode trap (SET) \cite{seidelin2006microfabricated}, such as detection \cite{todaro2021state, setzer2021fluorescence, reens2022high}, control electronics \cite{stuart2019chip} and optical beam delivery \cite{mehta2016integrated, niffenegger2020integrated, mehta2020integrated}. There has also been success miniaturizing the ultra-high vacuum chamber \cite{wilpers2013compact,aikyo2020vacuum} which houses the SET.
However, two of the most fundamentally limiting optical sub-systems today in terms of size, power, complexity, and reliability, are the precision low phase noise laser and the large ultra-low expansion optical cavity necessary to narrow the linewidth and stabilize the laser \cite{alnis2008subhertz}. 
These two subsystems constitute the stabilized laser, which is particularly important for optical clock and qubit operations, and even in compact trapped ion systems more than half of the volume is comprised of the laser and optics systems \cite{cao2017compact}. Further, these optical systems are extremely sensitive, requiring careful temperature control, vacuum chambers, and dedicated vibration isolation. The noise performance, stability, and reliability of the stabilized laser determines the achievable fidelity in accessing the ultra-narrow linewidth clock transitions and qubit operations and can limit quantum sensing performance. Today, atomic systems are pushing the fundamental limits of precision and becoming increasingly susceptible to high offset frequency laser phase noise due to techniques requiring short pulses which generate noise at high frequency offsets \cite{dayLimitsAtomicQubit2022}, creating crosstalk between atomic transitions and decreased stability through intermodulation distortion \cite{audoinLimitFrequencyStability1991a}. Further, the remote location of stabilized lasers and optics from the physics package requires noise stabilization techniques like fiber noise cancellation \cite{ma1994delivering} for active beam path stabilization that inhibit portable operation for quantum sensing.  
 
The most ubiquitous tool for atomic and molecular optical physics apparatus like trapped ion systems are the low noise visible light external cavity diode lasers (ECDLs) \cite{chen2022stable}, which require meter-scale rack and table-top setups for each precision wavelength needed to address each atomic transition. The standard method to perform optical clock and qubit experiments, as illustrated in Fig. \ref{fig:vision}a, is to lock the ECDL to an environmentally isolated optical reference cavity.
Such lasers provide low fundamental linewidth (FLW) through their large cavity mode volume and moderate integral linewidth (ILW) through external cavity stabilization. Recently, fiber Brillouin lasers, with their inherent nonlinear noise suppression and superior high-frequency laser noise properties, have been used to address the optical clock transition in a trapped ion \cite{loh2020operation}. But such fiber-based lasers operate in the mid-IR, requiring bulky, power consuming second harmonic generation for visible and near-IR clock transitions. Direct-drive visible light laser alternatives to the ECDL include self-injection locked (SIL) lasers based on fiber-coupled bulk-optic whispering gallery mode microresonators \cite{savchenkov2020application}. However, these lasers are bulky and do not provide further integration paths with other components of the optical infrastructure or with the surface electrode trap. Recently, promising progress has been made with ultra-low loss silicon nitride ($\text{Si}_3 \text{N}_4$) integration platform \cite{blumenthalSiliconNitrideSilicon2018} to realize ultra-low noise integrated visible and near-IR 674~nm and 780~nm stimulated Brillouin scattering (SBS) lasers \cite{chauhanVisibleLightPhotonic2021b, chauhanVisible780Nm2023} and SIL 780~nm lasers \cite{isichenkoChipScaleSubHzFundamental2023}. 

Laser pre-stabilization by locking to an optical reference cavity provides the low fractional frequency noise (FFN) required for locking to precision atomic transitions. Moving from table top reference cavities to the chip-scale is an essential step towards integrated stabilized lasers and monolithic integration with ion trap systems on-chip. There has been tremendous progress towards miniature bulk optic vacuum gap \cite{didier2018ultracompact, mclemoreMiniaturizingUltrastableElectromagnetic2022a} and solid silica cavities in ambient environment \cite{zhangMicrorodOpticalFrequency2019}. However, these technologies are not readily integrated on-chip with lasers and other components needed to create a fully integrated stabilized laser. Progress with on-chip waveguide reference cavities shows promising capabilities \cite{liu36HzIntegral2022, leeSpiralResonatorsOnchip2013} with a 4-meter silicon nitride coil \cite{liu36HzIntegral2022} stabilizing a semiconductor laser to a 36 Hz ILW ($1/\pi$) and an Allan deviation (ADEV) of $1.8\times 10^{-13} / \sqrt{\tau}$ at 10 ms measurement time with 2.3 kHz/s carrier drift. More recently, a 674~nm 3-meter long silicon nitride coil resonator reduced laser FN by over 4 orders of magnitude, yielding a 4.2 kHz ILW and $3.5\times 10^{-12} / \sqrt{\tau}$ ADEV at 20 ms \cite{chauhan2022integrated}. 

Progress towards implementing integrated stabilized laser technologies in an operational ion trap experiment has been limited. Such experiments must demonstrate measurements and operations critical to optical clocks, quantum computing, and sensing including ion spectroscopy, optical clock transition stability, qubit state preparation and measurement (SPAM) \cite{bruzewicz2019trapped}, and qubit coherence (e.g. Rabi oscillations and Ramsey interferometry). To date there has been limited progress in this direction using integrated photonic laser systems.

\begin{figure*}[]
\includegraphics[width=0.9\textwidth]{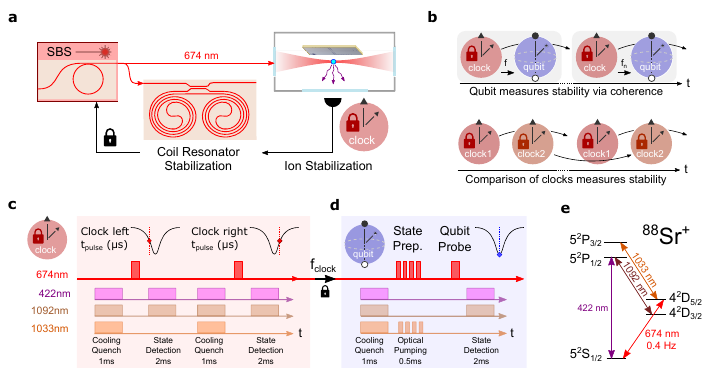}
\caption{
\textbf{Qubit operations using photonic coil stabilized laser: }
\textbf{a},
We demonstrate an integrated approach to generating and stabilizing the ultra-narrow linewidth laser. 
We use a photonic integrated direct-drive 674~nm ultra-low phase noise SBS laser, first stabilized to a photonic integrated 3-meter coil resonator reference cavity, then to a trapped ion optical clock transition for absolute stabilization. \textbf{b},
Optical clock operations are interleaved with trapped ion qubit operations to provide frequency stability for qubit operations, enabling us to measure the coherence of the laser with the qubit. We also interleave two independent optical clocks to measure the absolute frequency stability of the clocks by comparing their frequencies. 
\textbf{c},
Trapped ion optical clock laser pulse sequence to stabilize the laser, with one sample on the left full-width-half-max (FWHM) and one on the right side of the narrow optical clock transition ($S_{1/2} \rightarrow D_{5/2}$). Pulses of 422~nm light are used to cool the ion and to detect the ion state via fluorescence. 1092~nm light is used during cooling and detection to repump the ion out of the $D_{3/2}$ state. During cooling a 1033~nm laser is pulsed to repump the ion out of the $D_{5/2}$ state, `quenching' the optical qubit controlled by the 674~nm laser. 
\textbf{d},
Qubit operations are then interleaved with the clock to perform trapped ion qubit experiments with the frequency of 674~nm laser stabilized by the clock. 
The clock stabilized 674~nm laser is then used for qubit state preparation and qubit operations. 
\textbf{e},
Energy levels of the strontium trapped ion showing the optical clock transition at 674~nm ($S_{1/2} \rightarrow D_{5/2}$) which has a natural linewidth of 0.4 Hz.
}
\label{fig:architecture}
\end{figure*}

Here we report a transformational advance in photonic integration for trapped ion quantum and clock systems. We demonstrate for the first time, to the best of our knowledge, clock and qubit operations using chip-scale stabilized Brillouin laser, optical reference cavity, and ion trap technologies. The clock laser is a direct-drive ultra-low noise integrated \SiN 674~nm SBS laser stabilized to an integrated \SiN 3-meter long coil reference cavity which is then stabilized to the optical clock transition of a trapped strontium \Sr ion (Fig. \ref{fig:vision}b) in a surface electrode trap. We implement the same clock and qubit functions as performed in meter-scale experiments (Fig. \ref{fig:vision}a) with cm-scale technologies. We demonstrate for the first time the complete steps of trapped-ion spectroscopy, $99\%$ fidelity state preparation and measurement (SPAM), as well as Rabi oscillations and Ramsey coherence measurements, using direct drive integrated photonic technologies. We demonstrate a FLW of 12 Hz with the \SiN 674 nm SBS laser, which is a reduction of over 3500x from the free running pump FLW. With the SBS locked to the \SiN coil resonator, we demonstrate a reverse $1/\pi$ ILW of 580 Hz, which is a reduction of 300x from the the free-running pump reverse $1/\pi$ ILW of 180 kHz. This coil reference cavity is enabled by the lowest loss (0.66 dB/m) waveguides and highest quality factor (Q) resonator (93 million) to date at 674~nm. To precisely compensate for the coil drift, we adapt a clocking protocol to actively lock the laser to the ion transition itself, providing absolute frequency stability within 300 Hz. The SBS+coil+ion locked laser ADEV is measured to be 5$\times10^{-13} / \sqrt{\tau}$ fractional stability at 1 second with a frequency drift of $\pm$4 kHz over a minute.

We demonstrate the performance benefits of the chip-scale coil-ion stabilized SBS laser by using the optical clock driven qubit as a sensor to measure the closed-loop system coherence time to be 60 $\mu$s, with Ramsey fringe contrast decay. We also utilize the SBS laser stabilized to the coil and the 0.4 Hz quadrupole transition to perform precision ion-spectroscopy and measure 6 kHz linewidths, a reduction of 2x (12 kHz) for the coil-ion stabilized ECDL only. Equally important, this linewidth measurement is self-consistent with the qubit coherence time measurements. We also demonstrate that the reduction of high frequency noise by the SBS laser significantly improves the signal-to-noise ratio (SNR) and resolution of spectroscopy. We find that this combination of SBS nonlinear noise reduction and coil-resonator large mode-volume with low thermo-refractive noise (TRN) limit \cite{liu36HzIntegral2022,liuPhotonicCircuitsLaser2022,chauhanVisibleLightPhotonic2021b}, provides exquisite laser noise characteristics for ion spectroscopy and qubit coherence. 

These results demonstrate the potential for the CMOS foundry compatible \SiN ultra-low loss integrated photonic platform \cite{blumenthalSiliconNitrideSilicon2018} to realize stabilized direct-drive lasers that can be monolithically integrated into the trapped ion surface trap, and represent a key advance in the development of portable trapped-ion quantum sensors and clocks that are immune to vibrations and phase noise, and herald an era of portable quantum technologies.

\begin{figure*}[]
\includegraphics[width=0.97\textwidth]{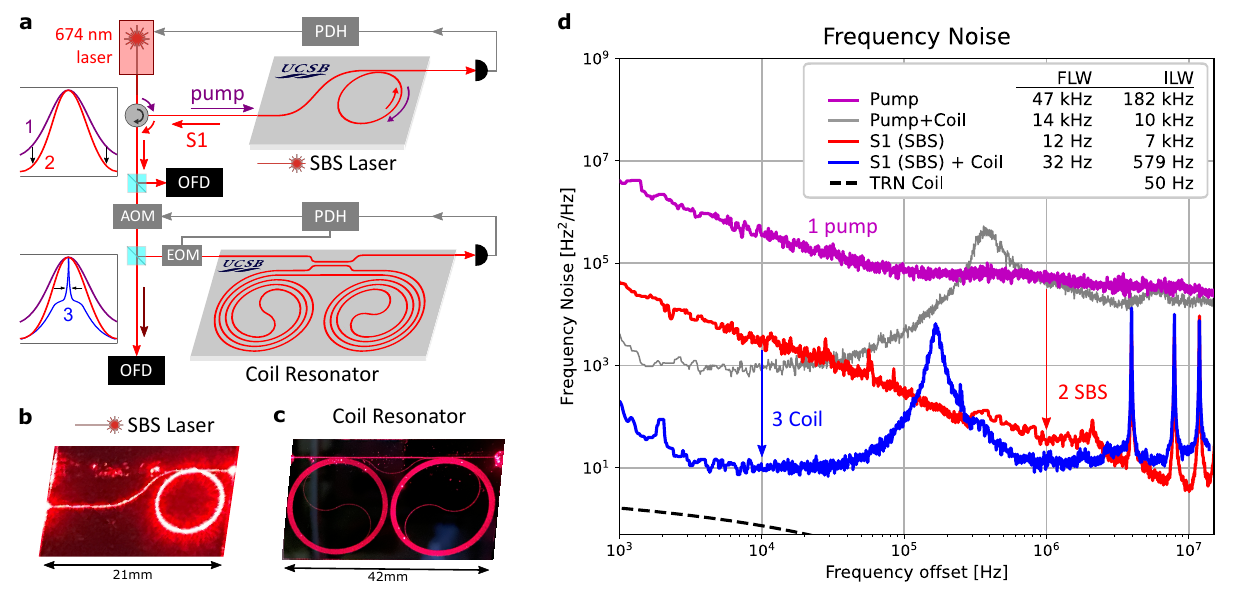}
\caption{
\textbf{Photonically stabilized narrow linewidth 674~nm laser:}
\textbf{a}, PDH lock of the SBS laser to the Integrated coil resonator further reduces the linewidth of the SBS laser and reduces drift.
\textbf{b}, Photograph of operational 674~nm integrated SBS laser. 
\textbf{c}, Photograph of operational 674~nm integrated coil resonator.
\textbf{d}, Frequency noise of each stage: free running pump laser (purple), SBS laser (without coil resonator lock) (red) and SBS laser locked to the coil resonator (blue). The FN of pump (without SBS) locked to just the coil resonator (grey) is also added for comparison. The ILW is calculated with $1/\pi$ method, integrated between 500 Hz and 330 kHz for the direct lock of pump to coil and between 500 Hz and 30 MHz for the rest of the FN. The SBS laser shows $>$ 3500x FLW reduction, reducing the ECDL pump laser FLW from 47 kHz to 12 Hz. The SBS laser also demonstrates an order of magnitude of ILW reduction, with ILW ($1/\pi$ method) reduced from 182 kHz to 7 kHz and with $\beta$ separation method reduced from 316 kHz to 31 kHz. The pump direct lock to the coil reduces the ILW by 18x to 10 kHz lock ($1/\pi$ method) and reduces the FN by 2-3 orders of magnitude. The PDH lock of SBS to the coil demonstrates an additional 2 orders of magnitude of FN reduction and the resulting coil stabilized SBS laser has a total ILW reduction of 300x to 580 Hz with $1/\pi$ method and 60x reduction to 6 kHz with $\beta$ separation method compared to the free running pump.
}
\label{fig:stabilizedlaser}
\end{figure*}

\section{Chip-Based Trapped-Ion Clock and Qubit Architecture}

The noise reduction and stabilization of the integrated laser is engineered in multiple stages. First, the SBS provides high frequency phase noise reduction (reducing the FLW of the pump laser by several orders of magnitude), then locking to the large mode-volume 3-meter coil resonator reduces the noise close to carrier (reducing the ILW). Lastly, we use the optical clock transition of a strontium ion trapped in a custom surface electrode trap to stabilize the laser absolutely via a clock protocol and counter any drift of the coil (Fig. \ref{fig:architecture}a). 
 
Next, we interleave the clock protocol (which continually stabilizes the laser frequency) with trapped ion experiments like qubit operations, Rabi oscillations or spectroscopy. We use interleaved qubit operations to characterize the laser noise via coherence measurements (Fig. \ref{fig:architecture}b). By interleaving the two independent clocks, we are able to compare their relative frequency to simultaneously measure the drift and the stability of the laser. 

The clock operations and timing diagram is shown in Fig. \ref{fig:architecture}c with 422~nm, 1092~nm, and 1033~nm lasers used for the trapped ion cooling, quench, and state detection cycles with the optical clock transition lock cycle measuring both sides of the transition for frequency lock to the ion. Once the coil-stabilized laser is locked to the optical clock transition (Fig. \ref{fig:architecture}c), the qubit operations (Fig. \ref{fig:architecture}d) start with state preparation and then probe the transition on resonance. The \Sr energy diagram and laser wavelengths are shown in Fig. \ref{fig:architecture}e. 

\begin{figure*}[]
\includegraphics[width=0.97\textwidth]{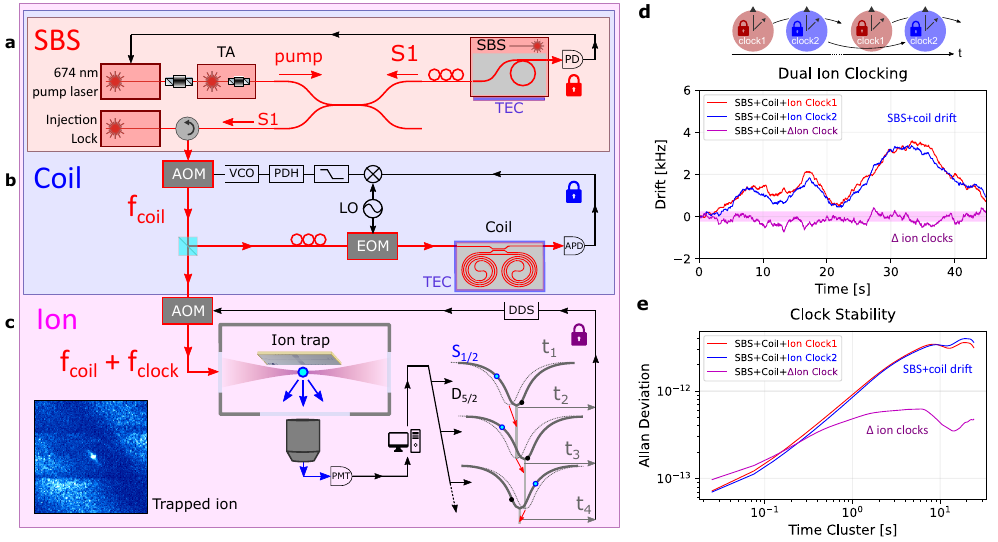}
\caption{ 
\textbf{Clock protocol for absolute laser stabilization using a trapped ion:}
\textbf{a}, In the first stage of integrated stabilization, the SBS laser system lowers the high frequency noise of the pump laser and reduces the ILW. 
\textbf{b}, Next, the SBS laser is stabilized through a PDH lock to a temperature controlled photonic 3-meter long coil-resonator. 
\textbf{c}, To further stabilize the laser, a clock protocol interrogates an optical clock transition of the trapped ion qubit at each FWHM and applies a correction to the frequency for the next loop. 
\textbf{d}, Running two independent clock feedback loops to the ion optical clock transition in parallel allows us to compare their performance and verify locking. Each lock (red and blue) shows the same underlying drift of the photonic coil resonator that must be corrected. The difference of the two locks (purple) shows the stability of the laser after locking to the optical clock transition. 
We also plot the RMS deviation of this frequency difference (purple box). 
\textbf{e}, Allan deviation of the coil stabilized SBS laser with and without ion clocking. 
}
\label{fig:clock}
\end{figure*}

\section{Integrated photonic coil resonator stabilized Brillouin Laser}
The integrated coil-stabilized laser consists of two \SiN chips, a 674~nm SBS laser and a 3-meter coil resonator (Fig. \ref{fig:stabilizedlaser}). The SBS provides nonlinear noise suppression of the high FN components and the coil-resonator provides suppression of close to carrier noise and reduction in fractional frequency noise (FFN) and drift \cite{liuPhotonicCircuitsLaser2022, liu36HzIntegral2022}. The high-Q integrated coil-resonator has a large mode volume which lowers the thermorefractive noise (TRN) floor, with TRN limited ILW of $< 50$ Hz, and provides a sharp discriminant slope for PDH lock at the \Sr clock wavelength of 674~nm \cite{chauhan2022integrated}. The coil resonator is fabricated in a CMOS foundry with the lowest propagation loss and highest Q demonstrated to date at 674~nm, with propagation loss of 0.65 dB/m, intrinsic quality factor ($Q_{i}$) of 93 million, and loaded quality factor $(Q_l)$ of 54 million, for the fundamental TM mode (See Supplementary Note 1 for further details of the waveguide and resonator design). The coil has a free spectral range of 65.5 MHz, meaning there are multiple peaks available in the AOM tuning range of 200 MHz enabling robust Pound-Drever-Hall (PDH) locking of the SBS laser to the coil resonator (Fig. \ref{fig:stabilizedlaser}a). Further details of the SBS laser and coil-resonator lock are provided in Supplementary Note 1. Photographs of the operational SBS laser and coil-resonator are shown in Fig. \ref{fig:stabilizedlaser}b and Fig. \ref{fig:stabilizedlaser}c respectively. The laser FN is measured using an optical fiber frequency discriminator \cite{liu36HzIntegral2022}. The resulting FN measurements are shown in Fig. \ref{fig:stabilizedlaser}d for the pump ECDL laser, SBS laser, SBS laser locked to the coil resonator using AOM, as well as direct lock of pump ECDL to the coil resonator using current modulation.  We demonstrate a FLW of 12 Hz with the SBS laser, which is a reduction of over 3500x from the 47 kHz free-running pump FLW.
With the SBS laser locked to the coil-resonator, we demonstrate a reverse $1/\pi$ ILW of 580 Hz, which is a reduction of 300X from the free-running pump reverse $1/\pi$ ILW of 180 kHz. The TRN limited noise floor is shown in Fig. \ref{fig:stabilizedlaser}d. The direct lock of the pump ECDL to the coil resonator is also performed for comparison, with ILW from $1/\pi$ integral method reducing from 182 kHz to 10 kHz, a reduction of 18x. To mitigate environmental effects, we packaged the coil resonator by attaching optical fibers and then placed it within an enclosure which we temperature stabilize to within a mK, details in Supplementary Note 4.

\section{Characterization of Photonic Stabilized Laser with Trapped Ion Qubit Quadrupole Transition}
\begin{figure*}[]
\includegraphics[width=0.97\textwidth]{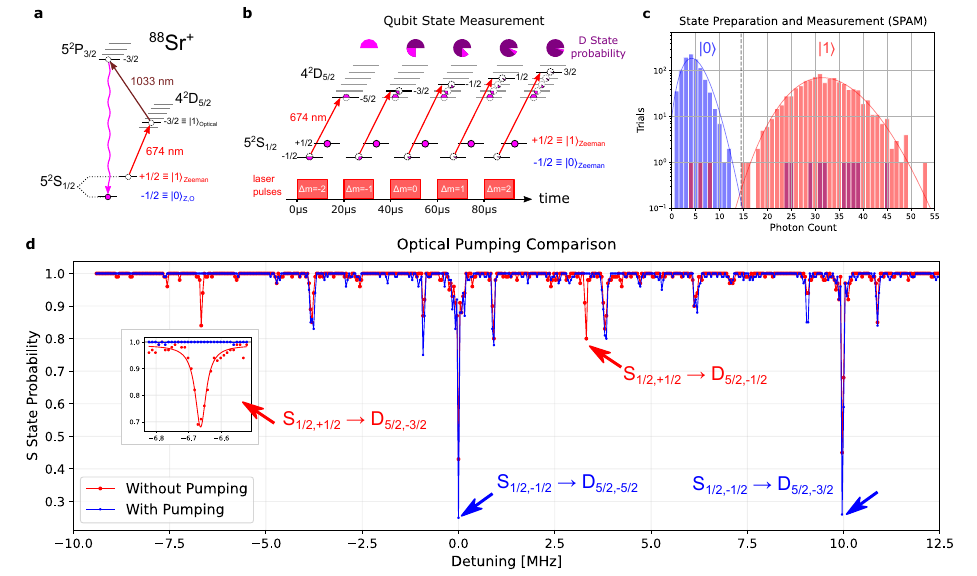}
\caption{
\textbf{Qubit operations for state preparation and measurement using photonic coil stabilized laser: }
\textbf{a}, Energy level diagram of frequency selective optical pumping scheme. First a 674~nm pulse excites any population in the $S_{1/2,+1/2}$ state to the $D_{5/2,-3/2}$ state, then a broad 1033~nm pulse pumps population to the $P_{3/2,-3/2}$ state, from which it decays to the $S_{1/2,-1/2}$ state. Each cycle of this sequence increases the probability the ion is in the $S_{1/2,-1/2}$ state.
\textbf{b},
For high fidelity state detection of a Zeeman qubit, one of the ground state spins must be shelved in the $D_{5/2}$ state before fluorescence detection. A single pulse of the 674~nm laser is capable of shelving most of the population but is limited by the coherence. 
Therefore, we apply pulses to all available D state sub-levels to ensure a high probability of shelving the desired Zeeman sub-level of the S state into the metastable D state. \textbf{c}, Histogram of photon counts over 1,000 trials showing a total SPAM fidelity of 99\%.
\textbf{d}, Ion spectroscopy with (blue) and without (red) optical pumping, shows that spectral lines initiating from the $S_{1/2,+1/2}$ states vanish with optical pumping. Optical pumping is performed on the $S_{1/2,+1/2} \rightarrow D_{5/2,-3/2}$ (at laser detuning f= -6.67 MHz) and the $S_{1/2,-1/2} \rightarrow D_{5/2,-5/2}$ (f=0 MHz) is used for qubit operations. The $S_{1/2,+1/2} \rightarrow D_{5/2,-1/2}$ (f= 3 MHz) transition also vanishes with optical pumping and the $S_{1/2,-1/2} \rightarrow D_{5/2,-3/2}$ (f= 10 MHz) remains.
Inset: a close up of the $S_{1/2,+1/2} \rightarrow D_{5/2,-3/2}$ shows over 99\% state preparation fidelity from the ratio of the amplitudes with and without optical pumping using only the coil+ion stabilized laser.
}
\label{fig:SPAM}
\end{figure*}

To characterize the photonically stabilized laser system, we use the narrow quadrupole transition ($S_{1/2} \leftrightarrow D_{5/2}$) of a strontium trapped ion qubit with a natural linewidth of 0.4 Hz.
The ion is trapped using an in-house fabricated surface electrode trap within a compact room temperature ultra-high-vacuum (UHV) chamber (see Supplementary Note 3). Broad spectroscopy scans show all expected transitions between the $S_{1/2}$ and $D_{5/2}$ sub-levels, which are Zeeman split by a 5.9~Gauss magnetic field, to avoid overlap and frequency cross-talk of nearby electronic and motional transitions (Fig. \ref{fig:SPAM}d).

\begin{figure*}[]
\includegraphics[width=0.97\textwidth]{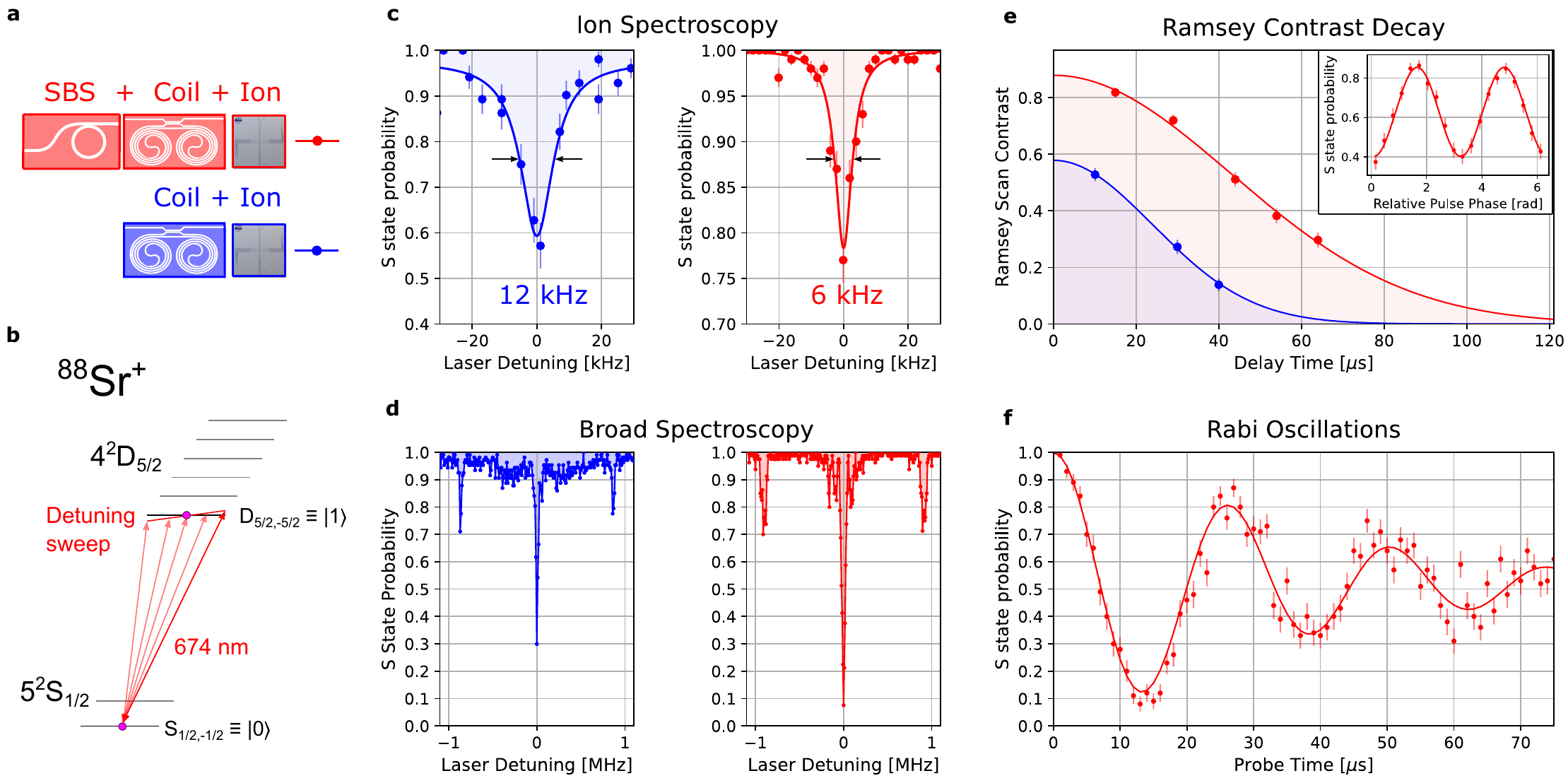}
\caption{
\textbf{Coherent qubit operations with a photonic and ion stabilized laser source:}
\textbf{a}, Comparison of the qubit operations and coherence of the coil stabilized pump laser while clocked to the ion (coil + ion, blue) and the coil stabilized SBS laser while clocked to the ion (SBS + coil + ion, red). 
\textbf{b}, Energy level diagram of the strontium trapped ion optical qubit. 
\textbf{c}, Trapped Ion spectroscopy with the coil+ion stabilized pump laser (blue) showing a linewidth of 12 kHz, while the SBS+coil+ion stabilized laser (red) shows a linewidth of 6 kHz.
\textbf{d}, Wider spectrum showing the motional sidebands of the trapped ion near $\pm$900 kHz. The SBS laser (red) suppresses the servo bumps that are otherwise seen in the coil+ion stabilized spectroscopy (blue). 
\textbf{e}, Ramsey contrast vs. delay time measuring the laser noise of the coil stabilized laser (blue) and the SBS laser stabilized to coil (red), showing an improvement from 33 $\mu s$ to 60.5 $\mu s$. Inset, shows the data for a single delay time as the phase between the two pulses is varied. 
\textbf{f}, Rabi oscillations with the SBS+coil+ion laser.
}
\label{fig:qubit}
\end{figure*}

To reduce the effect of thermal drift on skewing our spectroscopy data, we implement a novel ‘waterfall’ method of spectroscopy, in which each trial of the scan samples all frequencies in the scan once and then repeats for multiple trials to acquire the statistics to measure the probability of a transition. This is in contrast to a conventional ‘left-to-right’ spectroscopy scan which would sample each frequency a set number of trials from left to right and thus give different linewidths depending on the relative direction of the drift and the scan. By using this waterfall method we ensure the that linewidth can only be broadened, not artificially narrowed by the drift. We measured the linewidth of a single carrier electronic transition using this waterfall method and observe 12 kHz linewidth with just the coil + ion laser and 6 kHz with the SBS + coil + ion laser (Fig. \ref{fig:qubit}c). The bare coil was capable of ILW of 4 kHz \cite{chauhan2022integrated} and the SBS is capable of ILW of 600 Hz.

We first characterize the stability of just the photonic coil resonator before adding the SBS stage. To measure the coarse stability of the thermally stabilized photonic coil resonator we repeatedly performed spectroscopy scans and recorded the linewidth and center frequency. Typical drift of the coil over the course of an hour is less than 100 kHz. To measure the dependence of drift rate on temperature we tracked the coil drift after changing the temperature setpoint by 1 mK. We observe a coil frequency shift of $\sim$ 2.5 MHz in half an hour (1/e settling time of 7 minutes). This suggests that our observed frequency stability of 100 kHz over an hour is equivalent to a temperature stability of 30 $\mu K$.

\section{Ion Optical Clock stabilization}
To precisely track the drift of the coil and to simultaneously stabilize the laser to the absolute reference of the ion optical clock transition, we adapt a protocol similar to those used for optical clocks, Fig. \ref{fig:clock}b, which we implement via the software control (ARTIQ \cite{bourdeauducq_2016}) for our ion trap system. First we probe the left side of the ion transition at the full-width-half-max (FWHM) of the laser limited linewidth and record if the ion is bright or dark after fluorescence detection (2 ms). This indicates if the ion was excited to the D state by the laser pulse or not. Then we repeat on the other side of the resonance at FWHM and again record if the ion was bright or dark. If the laser is on resonance, both samples will have equal probabilities of excitation. However, if the laser has drifted, then the side to which the laser has drifted will be excited to the D state with higher probability. To correct this, the software lock will then apply a small correction to the frequency of the next cycle, with the feedback working to maintain balance between the excitation probabilities on each side of the ion resonance. 

In our case, the clocking protocol is relatively fast, as it is only limited by the detection time (2 ms) and cooling time (2 ms) for the ion between cycles, and not by the interrogation time ($\sim 60 \mu s$).
Locking the laser to the ion using this clocking protocol simultaneously maintains absolute frequency stability of the laser and precisely measures the laser drift (Fig. \ref{fig:architecture}a).
To verify robust clocking with the ion, we add a triangle wave to the frequency sent to the final AOM while clocking, which introduces an artificial \textit{drift} for which the clock protocol must provide feedback to adjust. Further, we run two independent locks in parallel, with only one having a triangle wave and one without, verifying the clock lock performance  accurately recovers the input triangle wave from the difference of the two clocks. This also verifies the lock is robust to artificial drift rates of more than 4 kHz/s.

Finally, to characterize the drift of the laser and the stability of the clock, we run two independent clocks without any artificial ramps and compare their stability. We observe that the SBS + coil + ion clock maintains an RMS frequency difference of 250 Hz from the ion transition (Fig. \ref{fig:clock}f), which is much smaller than the observed linewidth. Calculating the Allan deviation shows that the SBS + coil + ion has an fractional stability of $5 \times 10^{-13} / \sqrt{\tau}$ at 1 second (Fig. \ref{fig:clock}g). Following verification of clock lock and stabilization with the dual clocks technique, we interleave a single clock within each shot of qubit experiments (Fig. \ref{fig:architecture}c) to maintain the absolute frequency stability within 1 kHz of the transition. This final step enables us to reliably perform qubit operations on the ion with the combined ion and photonic stabilized lasers.

\section{Qubit state preparation and measurement (SPAM) with a photonic coil stabilized Brillouin laser}
Qubit state preparation and measurement are critical operations for quantum computation \cite{divincenzo2000physical} which are typically achieved with frequency selective narrow linewidth optical clock transitions. In this work we conclude that the performance of the coil + ion stabilized laser is sufficient for these operations, without the SBS laser. 
With only the pump ECDL locked to the 674~nm integrated coil resonator and stabilized with the ion, we perform high fidelity ($99\%$) qubit state preparation and measurement (SPAM) of a \Sr~trapped ion.
For state preparation we initialize the qubit in the $S_{1/2,-1/2}$ state via frequency selective optical pumping out of the $S_{1/2,+1/2}$ state. First, we pulse the 674~nm laser on the $S_{1/2,+1/2} \rightarrow D_{5/2,-3/2}$ transition for 15 $\mu s$. Next, we pulse a 1033~nm laser for 50 $\mu s$ to excite population from $D_{5/2,-3/2}$ to $P_{3/2,-3/2}$, which then rapidly decays to the $S_{1/2,-1/2}$ state (Fig. \ref{fig:SPAM}a). We then repeat this pulse sequence ten times. 
Fig. \ref{fig:SPAM}d shows two spectroscopy scans with (blue) and without (red) optical pumping operations, using the SBS + coil + ion laser. 
The absence of the $S_{1/2,+1/2} \rightarrow D_{5/2}$ transitions in the spectroscopy with optical pumping operations is the conclusive evidence of state preparation. 
The scans also show the depth of the $S_{1/2,-1/2} \rightarrow D_{5/2}$ transitions have increased, verifying state preparation. 
The inset of Fig. \ref{fig:SPAM}d shows the amplitude of the $S_{1/2,+1/2} \rightarrow D_{5/2,-3/2}$ transition with and without optical pumping, performed with just the coil + ion stabilized laser (no SBS), showing state preparation fidelity $>99\%$.

For state measurement we use the coil and ion stabilized laser to shelve one of the S state sub-levels into the D state (Fig. \ref{fig:SPAM}b). First, we apply a 15 $\mu s$ pulse on the $S_{1/2,-1/2} \rightarrow D_{5/2,-5/2}$ transition to shelve most of the $S_{1/2,-1/2}$. However, the fidelity of a single pulse is limited by the finite laser coherence time of the coil + ion laser. Therefore, we apply additional pulses to shelve the same S state sub-level into the other available D state sub-levels. Once we have shelved the $S_{1/2,-1/2}$ into the D states, we apply resonant 422~nm light on the $S \rightarrow P$ transition, along with 1092~nm light to repump the ion during cycling, and count photons with a photo multiplier tube for 2ms.
With just the coil+ion laser, we measure total state preparation and measurement (SPAM) fidelity $>99\%$. With the SBS+coil+ion laser, we observe that fewer pulses are necessary to achieve $>99\%$ fidelity (Fig. \ref{fig:SPAM}c), due to the increased coherence time, which makes each pulse more efficient in transferring population from the $S_{1/2,-1/2}$ state to the $D_{5/2}$ sub-levels. 

\section{Qubit Sensing for Measurement of Ion-Coil Stabilized Laser Coherence}

The quadrupole transition of \Sr has been used to create a universal qubit gate-set for quantum computation \cite{akerman2015universal}.
We find that the coil + ion stabilized laser alone is not sufficient for high fidelity coherent qubit operations, but that the full SBS + coil + ion laser system is more capable of high fidelity qubit operations due to its narrower linewidth, longer coherence time and reduced background of high frequency noise.

To measure the coherence of the lasers with the qubit we perform Ramsey interferometry on the trapped ion by applying two $\frac{\pi}{2}$ pulses to the ion ($S_{1/2} \leftrightarrow D_{5/2}$) while sweeping the relative phase of the pulses. This ensures we capture the maximum contrast at each delay time (Fig. \ref{fig:qubit}e, inset).
We then fit the decay of contrast of these Ramsey fringes as the delay time between the two $\frac{\pi}{2}$ pulses is increased. 
For the coil+ion stabilized laser, we observe a $1/e$ decay time of 33 $\mu s$ and for the SBS + coil + ion stabilized laser we observe 60.5 $\mu s$ (Fig. \ref{fig:qubit}e).
A coherence time of 60 $\mu s$ of the SBS + coil + ion stabilized laser suggests the laser is capable of linewidths of 5.3 kHz (eqn. \ref{eqn:coh}), which is similar to our observed linewidths of 6 kHz (Fig. \ref{fig:qubit}c). Our measurements of the coil + ion stabilized laser shows a broader linewidth of 12 kHz, which suggests a coherence time of 26 $\mu s$, similar to our measurement of 33 $\mu s$. These improvements to coherence highlight one of the main performance improvements yielded with the superior noise performance of the SBS laser. Another key measurement that highlights how the SBS + coil + ion spectroscopy has lower high frequency noise than the coil + ion is to compare broader spectroscopy scans (Fig. \ref{fig:qubit}d), which clearly show that the coil + ion has prominent servo bumps near to carrier (250 kHz) that are absent in the SBS + coil + ion spectrum.

\begin{equation}
\Delta \nu = \frac{1}{\pi \cdot \tau_{coh}}
\label{eqn:coh}
\end{equation}

Next, we perform Rabi oscillations on the trapped ion optical clock transition ($S_{1/2} \leftrightarrow D_{5/2}$) to test the fidelity of single qubit rotations using the coil stabilized SBS laser (Fig. \ref{fig:qubit}f).
We first prepare the ion with Doppler cooling (Fig. \ref{fig:vision}d), optical pumping (Fig. \ref{fig:SPAM}a,d), and then pulse the 674~nm laser for varying pulse lengths.
With the SBS + coil + ion stabilized laser we observe single flip fidelity of 92 \% limited by both the laser coherence time and the motional state of the ion. With the coil + ion stabilized laser we observe 80 \% contrast which decays rapidly due to the shorter coherence time. Still, this is sufficient for qubit SPAM operations as demonstrated above.

\section{Discussion}
In this work we present a transformational advance in integrated trapped ion quantum information processor, sensing, and timing systems. We report the first demonstration of photonic integration of the largest and most sensitive optical component of these quantum systems, namely the stabilized optical ion clock and qubit laser, and further lock this laser to the optical clock transition of a \Sr ion trapped on an integrated surface electrode chip. We then use this clock stabilized laser to perform interleaved qubit operations.
This set of results is achieved without vacuum isolation of the photonics and by leveraging relaxed temperature stability requirements of the integrated coil-cavity achieved through a clock stabilization protocol, dramatically lowering the technical overhead to achieve qubit operations (e.g. Rabi oscillations and Ramsey interferometry) and high fidelity SPAM with trapped ions.

Here, the chip-scale ion clock and qubit operations are performed with an integrated trapped ion chip and a photonic integrated silicon nitride direct-drive, 674~nm visible wavelength stimulated Brillouin laser which is stabilized to an integrated silicon nitride 3-meter coil resonator reference cavity. This level of performance is enabled by the exquisite noise properties of the Brillouin laser combined with the linewidth reduction and predictable drift of the low thermal-refractive noise (TRN) 3-meter coil-resonator cavity. The direct drive \SiN integrated 674~nm SBS laser produces 12 Hz FLW emission which represents over 3500X reduction in the ECDL pump laser FLW. Successive stabilization of the SBS laser to the \SiN integrated 674~nm 3-meter coil reference cavity further reduces the ILW of the free-running ECDL by $>$ 300x. 

The final stage of stabilization, locking the laser to a \Sr ion trapped on an integrated chip with  an optical clock protocol, results in record stabilization for an all-chip solution, providing absolute frequency stability within 300 Hz. We measure the long-term stability of the integrated coil-ion stabilized SBS laser to be 5$\times10^{-13} / \sqrt{\tau}$ at 1 second, with a frequency deviation of $\pm$ 4 kHz over a minute, ($\sim$100 Hz/second) which corresponds to temperature drift of less than 40 nK/s. The ion lock therefore effectively maintains temperature stability within 100 nK while clocking.

We demonstrate the performance of this approach by interleaving qubit operations within the ion clock stabilization cycles, using spectroscopy to measure a linewidth of 6 kHz and Ramsey interferometry to measure a self-consistent coherence time of 60 $\mu$s. We also perform qubit operations like Rabi oscillations and high fidelity SPAM ($>$99\%). We further compare the performance with and without the SBS laser to showcase how the high frequency noise reduction from the SBS laser dramatically improves the signal-to-noise ratio (SNR) and resolution of spectroscopy. 

These results demonstrate that precision, ultra-low noise chip-scale lasers can be stabilized and locked to atomic and qubit transitions with photonic integrated CMOS foundry compatible technologies without the aid of table-top or micro-optic reference cavities, leading to portable, reliable, and scalable trapped ion quantum systems. These results also show the potential for utilizing integrated technologies as pre-stabilization for the world's most stable lasers and highest precision atomic sensing experiments \cite{bothwellResolvingGravitationalRedshift2022}. 

These results are critical and direct steps on the path towards monolithic integration within a trapped ion quantum processor, as they demonstrate key constituent components which are all readily compatible with monolithic fabrication. Specifically, the ultra-low loss SiN platform \cite{blumenthalSiliconNitrideSilicon2018} used for fabrication of the photonics is CMOS compatible and therefore readily compatible with the ion trap chip fabrication. For example, the ion trap can be  directly fabricated on top of the cladding of the photonics. Monolithic integration is also motivated by the dramatically improved phase stability and vibration resilience provided by a monolithic beam path from the laser, through waveguides and grating couplers to a trapped ion \cite{niffenegger2020integrated}. This is in contrast to today's systems which must employ phase cancellation to mitigate the accumulation of noise in the free-space and optical fiber beam paths. This miniaturization would also enable the creation of arrays of portable quantum information processors for computation and sensing. Multiplexing of the ion clocks on-chip, as well as multiplexing of the laser stabilization devices on-chip, could create a path to much tighter phase stability clock protocols which interleave varying interrogations across various witness qubits.

Further, rapid advancement in photonic based laser stabilization performance in the infrared wavelengths such as cascading suppressed SBS \cite{liuIntegratedPhotonicMolecule2024a}, protocols \cite{lohUltranarrowLinewidthBrillouin2019, zhaoIntegratedReferenceCavity2021} and exploration of different materials to reduce drift \cite{zhaoLowlossLowThermooptic2020} suggests that there is still much room for improvement in the visible wavelengths which could enable even narrower visible light linewidths and lower phase noise, a subject of current ongoing research. Significant improvements are expected by direct tuning of the SBS cavity via thermal or PZT stress-optic actuation \cite{wangSiliconNitrideStressoptic2022} and on-chip modulation  to allow integration of the SBS and coil on a single chip while removing bulky and power consuming AOMs \cite{shaoIntegratedMicrowaveAcoustooptic2020}. Furthermore, development of heterogeneously integrated laser pump sources in the visible \cite{frankenHybridintegratedDiodeLaser2021, wenzelDistributedBraggReflector2022, isichenkoChipScaleSubHzFundamental2023} would enable locating the pump laser on chip and a fully integrated chip-scale photonic-ion platform. Combined with the vibration resilience of photonic delivery to the trapped ion qubit, this fully integrated chip-scale photonic-ion platform could open the door to ultra-precise \& portable, quantum sensors, optical clocks and quantum computers.

\section{Acknowledgments}
This work was supported in part by funding from ARO under award number W911NF2310179, ARL under award number W911NF2220056 and gift funding from Infleqtion and Quantinuum. The views, opinions and/or findings expressed are those of the author(s) and should not be interpreted as representing the official views or policies of the Department of Defense or the U.S. Government. The authors gratefully acknowledge help from Karl D. Nelson of Honeywell for chip fabrication. 

\section{Author Contributions}
R.J.N. and D.J.B. conceived of the work.
N.C. designed, packaged, characterized and analyzed the photonics with assistance from A.I. K.L. and R.J.N; 
J.W. fabricated the SBS devices; 
R.J.N. directed the ion experiments and analyzed the ion data.
C.C. fabricated the ion trap, assembled the UHV chamber and performed the ion experiments with assistance from N.H.;
All authors discussed the results and contributed to the writing of the paper. R.J.N. and D.J.B. supervised the research.

\section{Competing interests}
The Authors declare no competing interests.

\section{Data availability}
The data that support the plots
within this paper and other findings of this study are
available from the corresponding author upon reasonable
request.

\section{Code availability}
The codes that support the findings
of this study are available from the corresponding authors
upon reasonable request.

\section{Correspondence}
Correspondence and requests for materials should be addressed to R.J.N. (rniffenegger@umass.edu) and D.J.B. (danb@ucsb.edu).

\bibliography{ref}

\section{Supplemental Information}
In this Supplementary Information, we outline more details on the locking setups, design of the SBS and coil resonator and the enclosure to thermally stabilize the resonators.

\subsection{Supplementary Note 1: Details of lock}
SBS Lock: The 674~nm pump laser is a commercial external cavity diode laser (Moglabs Littrow cavity laser) which seeds a commercial tapered amplifier (TA). The pump lock to the SBS resonator is performed with the inbuilt proportional-integral-differential (PID) servo in the controller of the laser. The controller also modulates the current at 250 kHz to add sidebands for the lock. The laser comes with a 700 kHz PD for locking and the SBS resonator transmission is detected using this PD. Demodulation is carried out internally in the controller and a current servo is added with the drive current for the lock. This Pound-Drever-Hall (PDH) lock is a `weak' lock, i.e. it keeps the pump laser on resonance with the SBS resonator but does not provide any linewidth reduction.

SBS lock to coil: For locking SBS laser to the coil, the same APD, demodulation scheme and servo is used as the direct lock of the pump to coil with the difference being that an AOM is used frequency control and the servo signal is fed to the VCO controlling the AOM. For modulation, since the AOM response is limited, a resonant 25 MHz electro-optic modulator (EOM) is used.

\subsection{Supplementary Note 2: Resonator design}
\begin{figure*}[]
\includegraphics[width=0.9\textwidth]{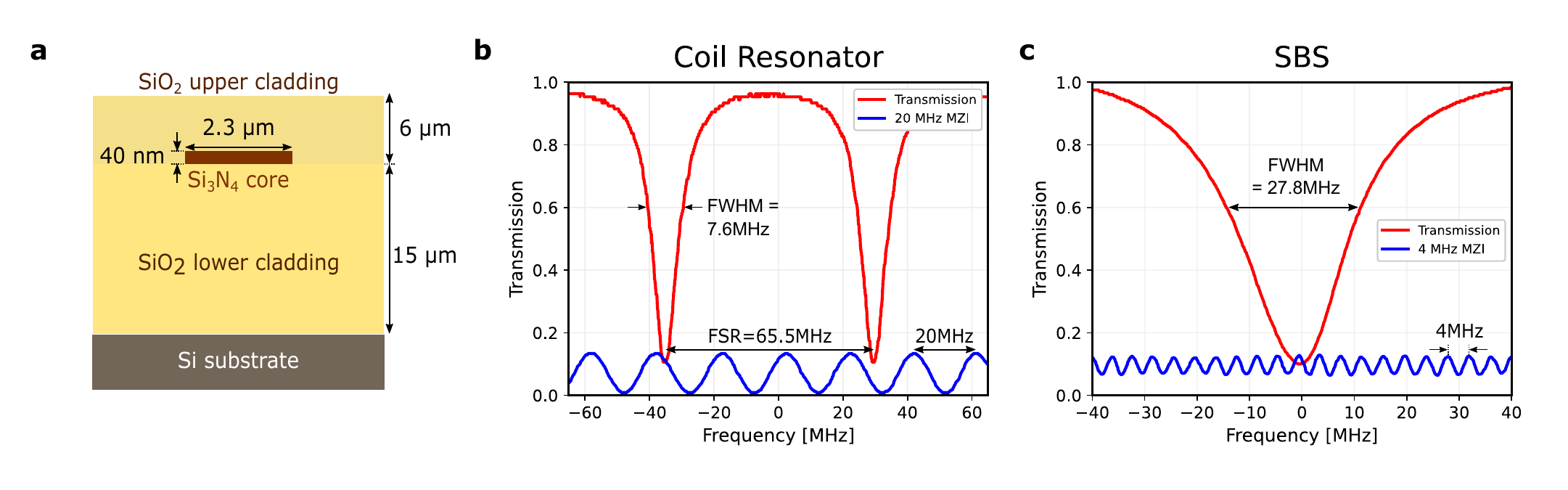}
\caption{
\textbf{Resonator design and characterization:}
(a) waveguide cross section. 
(b) Q and FSR measurements of the coil resonator with $Q_l$ of $54 \times 10 ^ 6$, $Q_{i}$ of $93.2 \times 10 ^ 6$ propagation loss of 0.66 dB/m and a measured FSR of 65.5 MHz . 
(c) Q and FSR measurements of the SBS resonator with $Q_l$ of $16 \times 10 ^ 6$, $Q_{i}$ of $39.6 \times 10 ^ 6$ propagation loss of 1.6 dB/m.
}
\label{fig:Qs}
\end{figure*}

The resonators are based on low loss dilute mode \SiN waveguides with a waveguide width of 2.3 $\mu$m and core thickness of 40 nm. The core is made of stoichiometric \SiN deposited with low pressure chemical vapor deposition process, the lower cladding is 15 $\mu$m thermal oxide and the upper cladding is 6 $\mu$m, formed with plasma enhanced chemical vapor deposition \cite{chauhanVisibleLightPhotonic2021b}. The waveguide supports the fundamental TE (TE0) and fundamental TM (TM0) modes and the cross section is shown in Fig. \ref{fig:Qs}a. The resonators are designed in the lower loss TM0 mode which also has a larger mode area of 2.4 $\mu$$m ^2$ compared to the TE mode area of 1.4 $\mu$$m ^2$. The critical bend radius, defined as radius above which bend loss contribution is $<$ 0.01 dB/m, is $\sim$ 3.5 mm for the TM0 mode. This waveguide design is used for SBS resonator and for the bus and coupling section of the coil resonator. The coil section of the coil resonator has waveguide width of 3.25 $\mu$m to reduce the bend losses and further increase the area of the mode to 3.1 $\mu$$m ^2$. The resonator coupling for both the coil and the SBS resonator is designed to only support the TM0 mode with critical coupling. The coil is processed in a 200 mm CMOS process. The Q and FSR of the coil obtained by calibrated MZI method \cite{chauhanVisibleLightPhotonic2021b,gundavarapuSubhertzFundamentalLinewidth2019}. The measured Q$ _l$ = 54 million, Q$ _i$ = 93.2 million with a propagation loss of 0.66 dB/m and FSR of 65.5 MHz is obtained, shown in Fig. \ref{fig:Qs}b, which is the lowest waveguide loss and highest Q at 674~nm. The SBS resonator are deigned so that both the pump and ~ 25 GHz redshifted first Stokes tone (Brillouin gain shift) are both resonant in the resonator \cite{chauhanVisibleLightPhotonic2021b}, and the FSR of 8.33 GHz is chosen such that 3x FSR = Brillouin gain shift. The SBS resonators are fabricated in a 100 mm wafer in UCSB cleanroom and demonstrates higher loss likely due to higher bend loss contribution resulting from fabrication variations increasing the critical bend radius, with propagation loss of 1.6 dB/m, $Q_l$ of $16 \times 10 ^ 6$ and $Q_{i}$ of $39.6 \times 10 ^ 6$, shown in Fig. \ref{fig:Qs}c. The SBS resonator demonstrates a threshold of 10 mW (on chip power) and a 40 mW S2 threshold. The SBS resonators are operated just below S2 threshold for lowest fundamental linewidth operation \cite{behuninFundamentalNoiseDynamics2018, gundavarapuSubhertzFundamentalLinewidth2019}. Future designs can improve the threshold by fabricating the CMOS devices also in the CMOS process.

\subsection{Supplementary Note 3: Ion trap fabrication}
The surface electrode ion trap chip was designed and fabricated in the UMass Amherst clean room facilities. The surface electrodes are composed of a 1.1 $\upmu$m thick layer of sputtered niobium metal, deposited onto a 4 in. fused silica wafer. The electrodes are defined via reactive ion etching (RIE) which transfers the optical lithography pattern to the niobium. The wafers are then diced into 1 cm square chips and cleaned with argon ion milling, which removes $\sim$ 100 nm of niobium and anneals the surface. 

\subsection{Supplementary Note 4: PIC packaging}
\begin{figure}[b]
\includegraphics[width=0.45\textwidth]{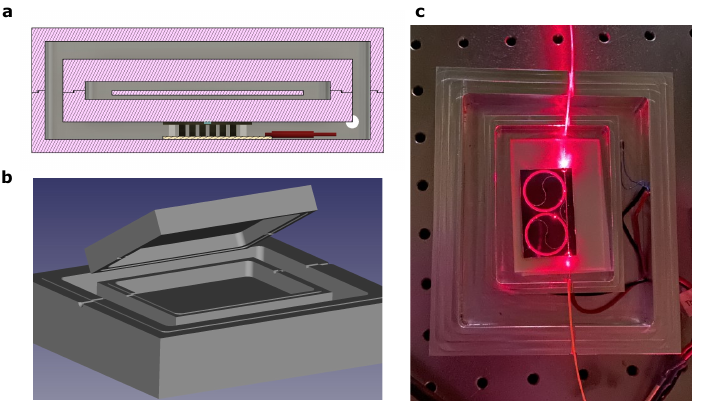}
\caption{ \textbf{Enclosure:}
(a) Side view of the dual aluminium enclosures for the packaged coil with a TEC in between.
(b) 3D view of the enclosures.
(c) Packaged coil on top of Teflon sheet (white) within enclosure.
}
\label{fig:enclosure}
\end{figure}

Temperature stabilized aluminium enclosures Fig. \ref{fig:enclosure}b house the integrated coil to isolate it from the environment. Cleaved fibers are aligned and epoxied to the facet of the coil, which is then placed upon Teflon Fig. \ref{fig:enclosure}c inside the inner aluminium enclosure. The inner enclosure is mounted on a TEC Fig. \ref{fig:enclosure}a and stabilized by a temperature controller (Vescent Slice-QTC) to provide the thermal control within a 1 mK. This aluminum box is placed inside an additional aluminum box and then a styrofoam box for further isolation. The SBS device is not packaged, but is temperature stabilized. The device is coupled with cleaved fibers using nano-positioners (Thorlabs Nanomax) and index matching gel is used to reduce facet loss. The facet loss for the packaged coil is ~ 5.5 dB/facet and is mostly expected from the misalignment during the packaging process, the facet loss for the SBS resonator is $\sim$ 2 dB/facet.

\end{document}